\documentclass[fleqn,10pt]{wlscirep}
\usepackage[utf8]{inputenc}
\usepackage[T1]{fontenc}
\title{Correcting model error bias in estimations of neuronal dynamics from time series observations}

\author[1]{Ian Williams}
\author[1]{Joseph D. Taylor}
\author[1,*]{Alain Nogaret}

\affil[1]{Department of Physics, University of Bath, Bath BA2 7AY, United Kingdom}

\affil[*]{A.R.Nogaret@bath.ac.uk}

\begin{abstract}
Neuron models built from experimental data have successfully  predicted observed voltage oscillations within and beyond training range.  A tantalising prospect is the possibility of estimating the unobserved dynamics of ion channels which is largely inaccessible to experiment, from membrane voltage recordings.  The main roadblock here is our lack of knowledge of the equations governing biological neurons which forces us to rely on surrogate models and parameter estimates biassed by model error.  Error correction algorithms are therefore needed to infer both observed and unobserved dynamics, and ultimately the actual parameters of a biological neuron.  Here we use a recurrent neural network to correct the outputs of a surrogate Hodgkin-Huxley (HH) model.  The reservoir-surrogate HH model hybrid was trained on the voltage oscillations of a reference HH model and its driving current waveform.  Out of the six reservoir-surrogate model architectures investigated, we identify one that most accurately recovers the reference membrane voltage and ion channel dynamics.  The reservoir was thus effective in correcting model error in an externally driven nonlinear oscillator and in reconstructing the dynamics of both observed and unobserved state variables from the reference model mimicking an actual neuron.
\end{abstract}
\begin{document}

\flushbottom
\maketitle
\thispagestyle{empty}

\section*{Introduction}

Building models from data is important for constructing digital twins of complex or chaotic dynamical systems such as neuroelectronic circuits~\cite{Brookings2014,Pescador2024,nogaret_approaches_2022} or the weather~\cite{Kalnay2003}, and inferring information hidden from experiment.  Takens' embedding theorem~\cite{Takens1981} underpins the idea that a model can be fully reconstructed from the information contained in observations of a dynamical system over a finite time interval~\cite{sprott_chaos_2003}.  As a result, a number of statistical~\cite{Lillacci2010,Lynch2015} and gradient descent methods~\cite{Toth2011,Waechter2005} have been developed that have successfully re-synchronized systems of nonlinear equations to time series data to estimate model parameters in biology~\cite{Brookings2014,Toth2011, nogaret_automatic_2016,mirams_optimal_2024,Morris2024}, chemistry~\cite{kravaris_advances_2013}, or model initial conditions in weather forecasting~\cite{Kalnay2003}.  A challenge specific to neuroscience is the lack of knowledge about the precise equations of neurons and networks~\cite{Wells2024,nogaret_automatic_2016}.  As a result, synchronizing a surrogate nonlinear model to data often yields multi-valued parameters that fail to represent the underlying biological processes they are meant to describe~\cite{OLeary2015,Gutenkunst2007}.  Training models to accurately describe the dynamics of underlying processes calls for correcting model error.  Model-free approaches~\cite{lukosevicius_reservoir_2009,Pathak2018} that rely on recurrent neural networks have provided an alternative approach to successfully predicting the oscillations of chaotic systems.  Recently Pathak et al.~\cite{pathak_hybrid_2018} have shown that embedding a model within a recurrent neural network can further extend the prediction window of chaotic oscillations.  The embedded model is used to generate an initial approximate prediction that the reservoir is trained to correct.  This hybrid computing approach successfully compensated for the error in the Lorenz oscillator model showing that it outperforms the accuracy of either the model or the reservoir used in isolation.  This analysis was extended to other chaotic oscillators by Duncan and R{\"a}th~\cite{duncan_optimizing_2023}, demonstrating that the hybrid approach performs well across a range of self-sustaining oscillators.

An important class of related systems are nonlinear oscillators driven by an external force.  For example, the oscillations of HH neurons~\cite{hodgkin_quantitative_1952} and networks are often driven by the current protocol of a current clamp or synaptic inputs.  It remains an open question whether hybrid approaches can effectively correct the errors of surrogate neuron models, and whether they are suitable for predicting the dynamics of both observed and unobserved state variables.  For example, the state of a biological neuron is described by the knowledge of its membrane voltage and the state of its ionic gates at any given time.  The membrane voltage however is the only variable that may be measured.  This means that information on the dynamics of ionic gates must be inferred indirectly from delayed measurements of the membrane voltage~\cite{Parlitz2014a,Letellier2005,Rey2014}.  The embedding procedure, which reconstructs the state vector from delayed voltage measurements, maximizes information transfer from data to model when both observability~\cite{Parlitz2014b} and identifiability~\cite{Schumann2016} criteria are satisfied and the dimension of the embedding space is at least twice the dimension of the state vector~\cite{Takens1981,Aeyels1981,Sauer1991}.  Applying model error correction in this setting therefore imposes additional constraints on the recurrent networks used for hybrid modelling, in particular their capacity to retain information over a sequence of delayed measurements greater than the dimension of the state vector.

Here we adapt the hybrid reservoir-model approach to correct error in surrogate models of a nonlinear neuronal oscillator driven by an external time dependent current.  During the training phase, the reservoir-HH model hybrid was stimulated both by the reference membrane voltage and the current protocol used to elicit this voltage while the output weights were optimized.  During the prediction phase, the reservoir-HH model was stimulated by the current protocol only to predict the membrane voltage oscillations within and beyond the training range.  We investigated six reservoir-HH model architectures and four types of surrogate models incorporating deliberate error in their parameters to determine the best error correcting architecture.  The results were validated by comparing the prediction accuracy of hybrid architectures to those of the surrogate model, and the reference data.  The best reservoir-HH model architecture was found to successfully recover both the membrane voltage oscillations and the unobserved gate dynamics of the reference model.  Predictions were also remarkably stable in response to rapid changes in frequency or amplitude of the forcing current.  In contrast predictions of the stand-alone reservoir were often driven to saturation by fast or abrupt changes in the current drive.  We discuss the contradictory expectations of a long reservoir retention time to correct error in the unobserved gate variables, against the need for a short reservoir retention time to respond to fast stimulation.  The solution of combining the reservoir with a surrogate model proved effective in using their complementary kinetic.  In addition, the hybrid architecture was very effective in correcting model error when multiple parameters were detuned by four-orders of magnitude from their reference value.  At vanishingly small model error, the prediction accuracy of hybrid architectures was however limited by intrinsic reservoir noise.  The hybrid approach demonstrates the ability to infer the complete dynamic state of a neuron from approximate conductance-based models.

\section*{Methods}

\subsection*{Neuron models}

We consider a reference model of the form $d\mathbf{x}/dt=\mathbf{F}(\mathbf{x}(t),\mathbf{p},I(t))$ where $\mathbf{x}(t)$ is the vector representing the neuron state at instant $t$, $\mathbf{p}$ is the set of model parameters (see Methods), and $I(t)$ is the current protocol driving the neuron oscillations.   The state vector $\mathbf{x}(t)$ has four components: the neuron membrane voltage $V(t)$ which is observable, and the gate variables of the Hodgkin-Huxley model $m(t)$, $h(t)$ and $n(t)$~\cite{Hodgkin1952} which are unobservable.  This model was used to generate a reference voltage time series $V_{mem}(t)$ that will be used to train recurrent networks in recovering the dynamics of the reference model.  In general, the reference $V_{mem}(t)$ time series will be measured from a biological neuron whose model equations are unknown.  Hence one must assume that this time series is fitted with a surrogate model $\mathbf{G}()$.  In this paper, we consider surrogate models obtained by detuning one or more parameters from their reference value $\mathbf{p}\rightarrow \mathbf{p}_{\epsilon}$ hence $\mathbf{G}()\equiv \mathbf{F}(\mathbf{x}(t),\mathbf{p}_{\epsilon},I(t))$.  Both reference and surrogate model have the following equation yielding the membrane voltage:

\begin{equation}
C\frac{dV}{dt}= g_{Na}m^3h(E_{Na}-V)+g_Kn^4(E_K-V)+g_L(E_L-V))+I(t), \\
\label{eq:eq1}
\end{equation}

\noindent and equations for the gate variables:

\begin{equation}
\frac{dg}{dt}= \frac{g_{\infty}(V)-g}{\tau_{g}(V)},\;\;\;\;\; g \equiv \{m,n,h\} \\
\label{eq:eq2}
\end{equation}

\noindent where $m$ is the Na activation gate, $h$ is the Na inactivation gate, $n$ is the K activation gate, $C$ is the membrane capacitance; $g_{Na}$, $g_K$ and $g_{Leak}$ are the areal conductance of the Na, K and Leak ion channels.  $E_{Na}$ and $E_K$ are the reversal potentials.  In the steady state, the gate activation have a sigmoidal dependence on $V$:

\begin{equation}
g_{\infty}(V)= 0.5 \left\{1 + \tanh \left[(V(t) - V_{g})/dV_g\right] \right\}, \\
\label{eq:eq3}
\end{equation}

\noindent and the gate recovery times are:

\begin{equation}
\tau_{g}(V) = \tau_{0,g} + \epsilon_g \left\{ 1 - \tanh^2 \left[(V(t) - V_{g}) / dVt_g \right] \right\},
\label{eq:eq4}
\end{equation}

\noindent where $\tau_{0,g}$ and $\tau_{0,g}+\epsilon_g$ are the minimum and maximum relaxation times, $dV_g$ is the width of the transition from gate open to gate closed, and $dVt_g$ is the width of the Bell-shaped voltage dependence of the recovery time.

\begin{figure*}
\centering
\includegraphics[width=\linewidth]{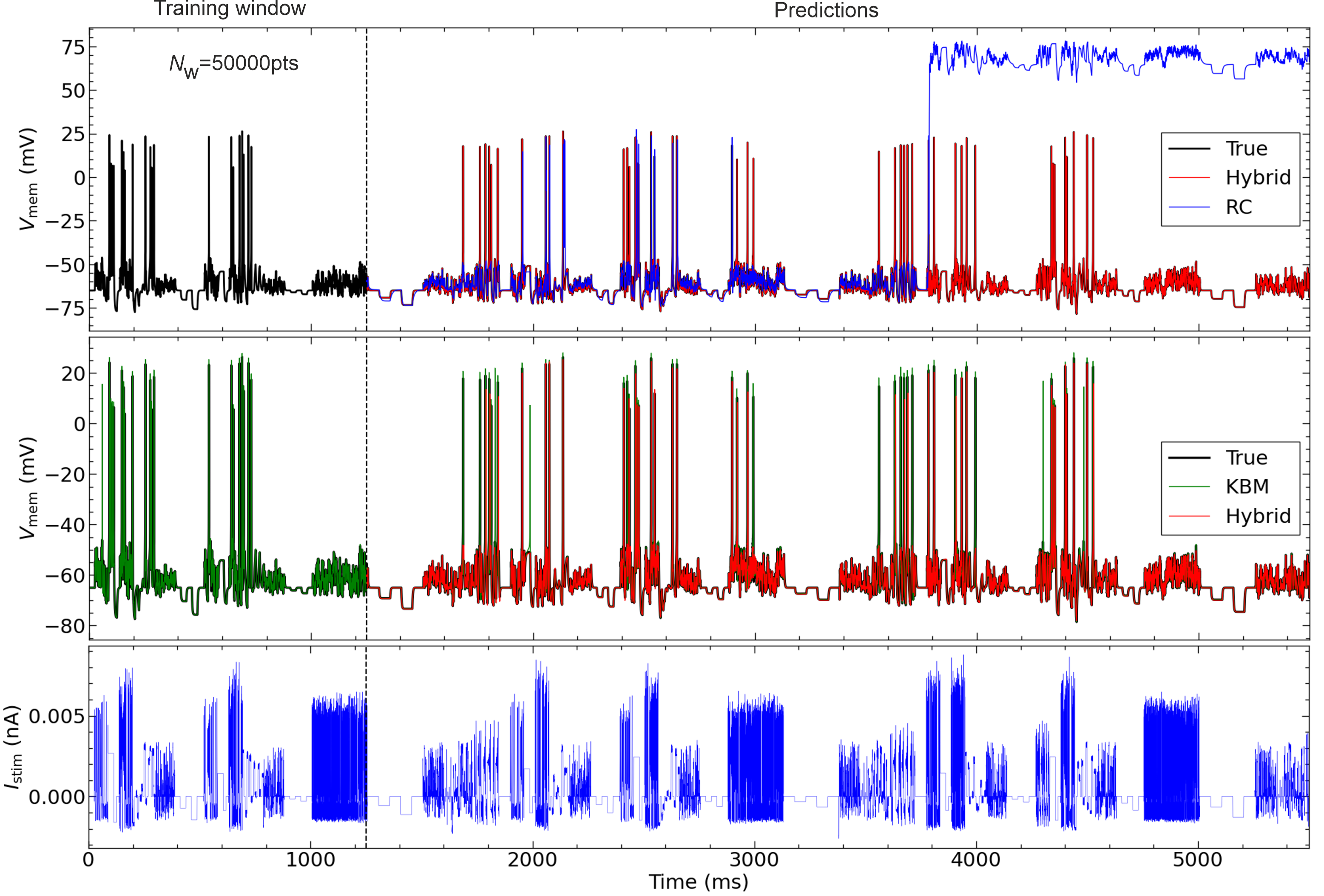}
\caption{\textbf{Predictions of the Reservoir and Reservoir-HH model hybrid} \\ Reservoir and Reservoir-HH model hybrids are trained on the time series of reference $V_{mem}(t)$ (black line) while driven by current protocol $I(t)$ (blue line).  The training window, 0-1250ms, has $N_W=5\times10^4$ data points.  From 1250ms onwards, the trained Reservoir (blue trace) and Reservoir-HH models (red traces) are driven by $I(t)$ in input and output the predicted neuron voltage.  $V_{mem}(t)$ (black line) is the reference.  (a) Reservoir and Reservoir-HH model predictions.  The embedded HH model incorporates no model error.  (b) Surrogate HH-model (KBM) and Reservoir-surrogate HH model (Hybrid) incorporating model error ($\epsilon_g=10\%$) in the surrogate HH model.  (c) Current waveform.  Sampling interval: $\Delta t = 25\mu$s.}
\label{fig:fig1}
\end{figure*}

The reference voltage times series $V_{mem}(t)$ was computed by integrating a current protocol $I(t)$ with a fifth order adaptive step-size Runge-Kutta Cash-Karp method (Fig.~\ref{fig:fig1}).  The current waveform $I(t)$ (blue trace) was chosen to elicit the maximum amount of information from the reference HH model so that the $V_{mem}(t)$ time series (black trace) fulfills the identifiability criterion~\cite{Wells2024}.  It incorporated a combination of current steps and chaotic oscillations of varying amplitudes and durations.

\subsection*{Surrogate neuron models}

Model error was simulated by detuning three parameters from their reference value:  the sodium channel conductance $g_{Na} \rightarrow g_{Na}(1+\epsilon_g)$, the sodium activation threshold $V_{m} \rightarrow V_{m}(1+\epsilon_v)$, and the recovery time of sodium activation $\tau_m \rightarrow\tau_m(1+\epsilon_\tau)$. $\epsilon_g$ represents an error producing a \textit{linear} dependence in the membrane voltage, whereas $\epsilon_V$ and $\epsilon_\tau$ represent error producing a \textit{nonlinear} change in the membrane voltage.

\subsection*{Reservoir architecture for a driven neuronal oscillator}

\begin{figure}
\centering
\includegraphics[width=\linewidth]{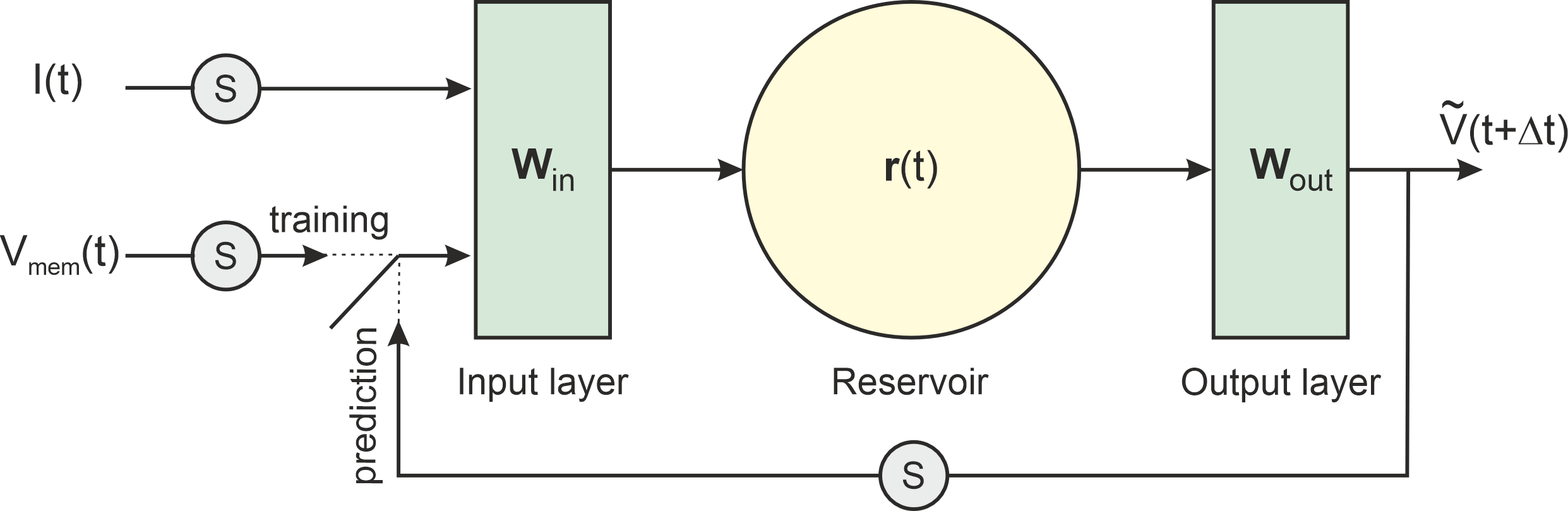}
\caption{\textbf{Reservoir for a current driven neuron} \\ In training mode (switch horizontal), the input layer ($W_{in}$) connects half of reservoir nodes to $V_{mem}(t)$ the other half to $I(t)$.  The weights of the output layer ($W_{out}$) are calculated to match the output voltage $\tilde{V}(t+\Delta t)$ to $V_{mem}(t+\Delta t)$.  In prediction mode (switch vertical), $\tilde{V}(t+\Delta t)$ is fed back into the input layer in place of $V_{mem}(t+\Delta t)$.  The reservoir had $N_R=1000$ nodes.}
\label{fig:fig2}
\end{figure}

We first constructed a stand-alone reservoir architecture (Fig.\ref{fig:fig2}) to model a current driven neuron.  This implementation of reservoir computing is novel as earlier work focussed on nonlinear oscillators with no driving force applied~\cite{pathak_hybrid_2018,duncan_optimizing_2023,Burghi2025,Pathak2018}.  Our reservoir architecture is adapted from Pathak et al.~\cite{pathak_hybrid_2018} by adding a second input $I(t)$ to the $V_{mem}(t)$ time series updating the network.  For optimum performance, the mean of both time series was re-centered on zero over the training window, and their oscillation range was re-scaled to the $[-\sigma, +\sigma]$ interval ($\sigma<1$).  This was done by means of scaling function $S$() (Fig.\ref{fig:fig2}) defined as:

\begin{eqnarray}
\left\{
\begin{array}{ccc}
S(V_{mem}) & = & \sigma\frac{V_{mem}-\langle V_{mem} \rangle}{\sqrt{\langle (V_{mem} - \langle V_{mem} \rangle)^2 \rangle}}, \\
S(I) & = & \sigma\frac{I-\langle I \rangle}{\sqrt{\langle (I - \langle I \rangle)^2 \rangle}},
\end{array}
\right.
\label{eq:eq5}
\end{eqnarray}

\noindent At every time step, half the $N_R$ reservoir nodes are updated by $S(V_{mem})$ and the other half by $S(I)$ through input matrix $\mathbf{W}_{in}$ given by:

\begin{eqnarray}
\mathbf{W}^T_{in}=\left(
\begin{array}{cccccc}
0 & \dots & 0 \; ,& 1 & \dots & 1 \\
1 & \dots & 1 \; ,& 0 & \dots & 0
\end{array}
\right).
\label{eq:eq6}
\end{eqnarray}

\noindent Each reservoir node is also updated by other reservoir nodes to connected to it.  The reservoir connectivity is determined by adjacency matrix $\mathbf{A}$.  The state of the reservoir is then updated though sum and thresholding as:

\begin{equation}
\mathbf{r}(t + \Delta t) = \tanh \left[ \mathbf{A} \mathbf{r}(t) + \mathbf{W}_{\mathrm{in}}\mathbf{v}_{\mathrm{in}}(t) \right],
\label{eq:eq7}
\end{equation}

\noindent where $v_{\mathrm{in}}\equiv\left\{S(V_{mem}),S(I)\right\}$ is the input vector.  The adjacency matrix was a symmetrical (undirected) array of 1s and 0s.  Connections between nodes were random and represented by 1s.  When modelling neurons with the stand-alone reservoir of Fig.\ref{fig:fig2}, we used an adjacency matrix with rank $N_R=1000$, spectral radius $\rho=1.25$, degree $D=6$ and input scaling factor $\sigma=0.4$.  $D$ specified the fraction of non-zero matrix elements reaching a node which is this case is $(D/N_R)=0.6\%$.  Prior to each training run, the reservoir state was initialised to zero by setting $\mathbf{r}(t=0)=(0,\dots,0)^T$ in Eq.\ref{eq:eq7}.

The reservoir architecture outputs $\widetilde{V}(t+\Delta t)$ as a linear superposition of the $N_R$ reservoir outputs $\mathbf{r}(t+\Delta t)$ weighted by the matrix elements $\mathbf{W}_{out}$ as:

\begin{eqnarray}
\left \{
\begin{array}{ccc}
\widetilde{V}(t) & = & \mathbf{W_{\mathrm{out}}}\mathbf{r}(t) + \mathbf{w}_{out},  \\
\mathbf{w}_{out} & = & \langle \mathbf{V}_{mem}(t) \rangle - \mathbf{W}_{out}\langle \mathbf{r}(t) \rangle.
\end{array}
\right.
\label{eq:eq8}
\end{eqnarray}

\noindent During the training phase, the output weights were calculated by matching the sequence of $(N_{W}-1)N_R$ reservoir states $\mathbf{R} \equiv \left\{\mathbf{r}(\Delta t),\dots,\mathbf{r}(N_W \Delta t)\right\}$ to the $N_{W}-1$ sequence of reference voltages $\mathbf{V}_{mem} \equiv \left\{ V_{mem}(\Delta t), \dots , V_{mem}(N_W\Delta t) \right\}$ across the 1250ms long training window (Fig.\ref{fig:fig1}).  This window had $N_W=50001$ data points equally spaced by $\Delta t=25\mu$s.  Once both sequences were re-scaled as $\mathbf{H} \equiv S(\mathbf{R})$, $\mathbf{Y} \equiv S(\mathbf{V}_{mem})$, the output weights were calculated through ridge regression using a Tikhonov regularisation~\cite{n_stability_1943} constant $\beta=10^{-4}$:

\begin{equation}
\mathbf{W}_{out} = \frac{\mathbf{YH}^T}{\mathbf{HH}^T+\beta \mathbf{I}}.
\label{eq:eq9}
\end{equation}

\noindent Once the weights have converged, training is complete.  The switch closing the voltage loop moves to the vertical position (Fig.\ref{fig:fig2}).  The reservoir is then set for predicting the voltage beyond the training window in response to current waveform $I(t)$ (Fig.\ref{fig:fig1}(a), blue trace).  We verified that a training window with $50001$ points was sufficient for the weights to converge.  Increasing $N_W$ to $60001$ or $70001$ data points did not noticeably change weight estimates.

\subsection*{Reservoir - HH model architectures}

\begin{figure*}
\centering
\includegraphics[width=\linewidth]{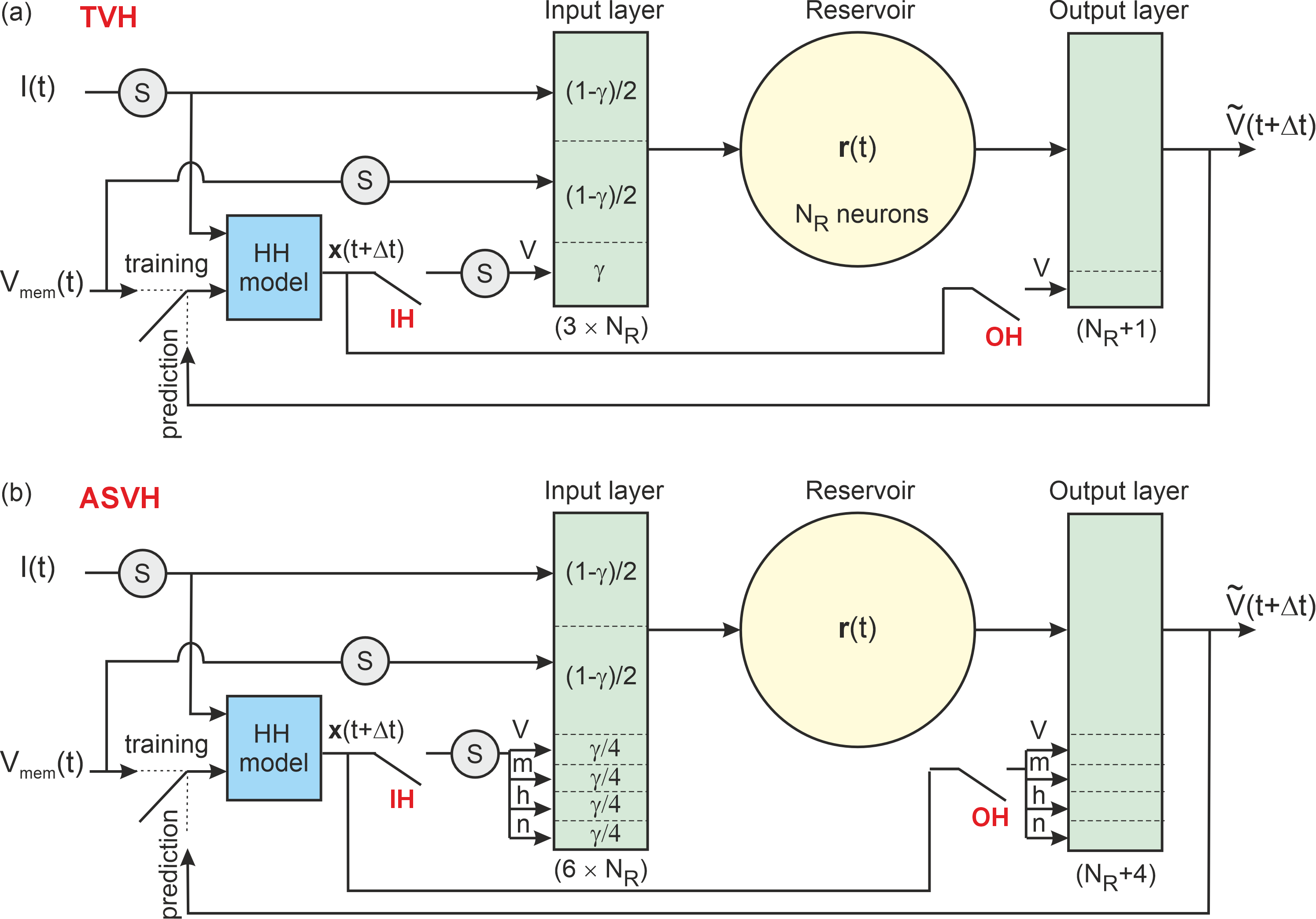}
\caption{\textbf{Reservoir-HH model architectures for model error correction} \\
(a) Target Variable Hybrid (TVH) architecture.  The surrogate HH model computes the state vector $\mathbf{x}$ at the next time step.  It only passes the membrane voltage vector component $V$ to the input/output layers of the reservoir.  (b) All State Variable Hybrid architecture (ASVH).  All 4 components of state vector $\mathbf{x}$: $V$, $m$, $h$, $n$ are now passed to the ipnput/output layers of the reservoir.}
\label{fig:fig3}
\end{figure*}

In order to correct the error bias of the surrogate model on the state vector when fitting $V_{mem}(t)$, we coupled the model to a reservoir.  The premise is that as model error increases, the reservoir will increasingly compensate for model error bias.  The hypothesis to test is thus whether the trained reservoir-surrogate model system can recover the dynamics of the reference model by predicting the oscillations of observed and unobserved state variables within and beyond the training range.  The dynamics of unobserved gate variables is implicitly contained in time delayed observations of $V_{mem}$~\cite{Rey2014,Schumann2016,Sauer1991}.  To this end, we investigated two reservoir-HH model architectures in Fig.\ref{fig:fig3}.  These extend earlier approaches~\cite{pathak_hybrid_2018,duncan_optimizing_2023}, by adapting them to a dynamically driven nonlinear oscillator and by applying error correction to unobserved state variables.

\begin{table}[t]
\begin{center}
\begin{tabular}{|l | c c c c c |}
\hline
\hline
\;\;\; Hyperparameter \;\;\; & \;\;\; $\rho$ \;\;\;& \;\;\; $D$ \;\;\; & \;\;\; $\sigma$ \;\;\; & \;\;\; $\beta$ \;\;\; & $N_R$ \;\;\; \\
\hline
\;\;\; Reservoir \;\;\;& 1.25 & 6 & 0.4 & $10^{-4}$ & $10^3$ \\
\;\;\; Reservoir-HH model \;\;\;& 1 & 8 & 0.8 & $10^{-3}$ & $10^3$ \\
\hline
\hline
\end{tabular}
\end{center}
\caption{\textbf{Hyperparameters used in Reservoir and Reservoir-HH architectures} Spectral radius $\rho$, degree $D$ of the adjacency matrix, input scaling $\sigma$, regularization term $\beta$, number of reservoir nodes $N_R$.}
\label{tab:tab1}
\end{table}

Both architectures embed the surrogate HH model before the input layer.  The HH model is updated with $I(t)$ and $V_{mem}(t)$ at time $t$ and computes the state vector $\mathbf{x}(t+\Delta t)$ at the next time step.  The integration of Eqs.\ref{eq:eq1}-\ref{eq:eq4} is done with the Cash-Karp algorithm.  The input layer is then updated with the model output, $I(t)$ and $V_{mem}(t)$.  The first architecture, Target Variable Hybrid (TVH), only passes the voltage state variable to the input layer (Fig.\ref{fig:fig3}a).  The input vector is then $\mathbf{v}_{\mathrm{in}}\equiv$($S(V_{mem})$,$S(I)$,$V$).  The second architecture, labelled All State Variable Hybrid (ASVH), passes all four state variables to the input layer (Fig.\ref{fig:fig3}b).  In this case the input vector is $\mathbf{v}_{in}\equiv$($S(V_{mem})$,$S(I)$,$V$,$m$,$h$,$n$).

We investigated variants of the TVH and ASVH architectures injecting the model output in the input layer only [Input Hybrid (IH)], the output layer only [Output Hybrid (OH)] or both [Full Hybrid (FH)] (Fig.\ref{fig:fig3}).  Also investigated was the effect of varying the fraction of reservoir nodes $\gamma \in [0,1]$ updated by the model variables.  In the TVH architecture, $\mathbf{W}_{\mathrm{in}}$ is a $3 \times N_R$ dimensional matrix connecting $(1-\gamma)N_R/2$ nodes to $I$, $(1-\gamma)N_R/2$ nodes to $V_{mem}$, and $\gamma N_R$ nodes to $V$.  The matrix of output weights, $\mathbf{W}_{\mathrm{out}}$ had one additional vector component weighting in model output $V$ in the calculation of $\widetilde{V}$.  In the ASVH architecture, the input layer $\mathbf{W}_{in}$ is a $6 \times N_R$ dimensional matrix connecting $(1-\gamma)N_R/2$ nodes to $I$, and $V_{mem}$ each, and $\gamma N_R/4$ nodes to $V$, $m$, $h$ and $n$ each.  Similarly, the output layer, had four additional weights for $V$, $m$, $h$ and $n$ when connected to the model.

\section*{Results}

\subsection*{Comparison Reservoir-only and Reservoir-HH model hybrid}

Fig.\ref{fig:fig1}a compares the membrane voltage predicted by the reservoir (blue trace) and the ASVH-FH architecture (red trace) to the reference (black trace).  Both systems were trained over the first 1250ms of data. Predictions were generated by integrating the current protocol of Fig.\ref{fig:fig1}c from 1250ms onwards.  The stand-alone reservoir is found to adequately predict sub-threshold oscillations but misses some of the action potentials seen in the reference time series (black line).  Reservoir predictions are also unstable.  The abrupt saturation of the reservoir output observed at $t=3800$ms is triggered here by the sudden increase in amplitude of the current drive at that point.  The Hodgkin-Huxley model does not exhibit chaotic behaviour under aperiodic forcing, hence we used the Root Mean Square Expectation (RMSE) to quantify the deviations of predictions from $V_{mem}(t)$ over a 1500 points window.  The prediction of the ASVL-FH architecture embedding the reference HH model (red trace) predicts both subthreshold and action potentials to a very high degree of accuracy.  The result validates the hybrid approach in extending the prediction range of the stand-alone reservoir and in showing that reservoir noise, although finite as we shall see below, has no detrimental impact on prediction accuracy.

Fig.\ref{fig:fig1}b compares the predictions of an erroneous HH model (green trace) and the same model corrected by the ASVH-FH architecture (red trace) to the reference voltage (black trace).  In this example, the surrogate HH model used a sodium conductance $g_{Na}$ detuned by $\epsilon_g=10\%$ from its reference value (see Methods).  Model error is seen to induce extra action potentials at 1000ms, 2000ms and generally to delay action potentials while increasing their amplitude.  The ASVL-FH corrected trace is seen to correct all discrepancies, restoring the predictions of the reference model.  The reservoir thus successfully corrects model error in the observed membrane voltage beyond the training window.

\subsection*{ASVH and TSV Reservoir-HH model architectures}

\begin{figure*}
\centering
\includegraphics[width=\linewidth]{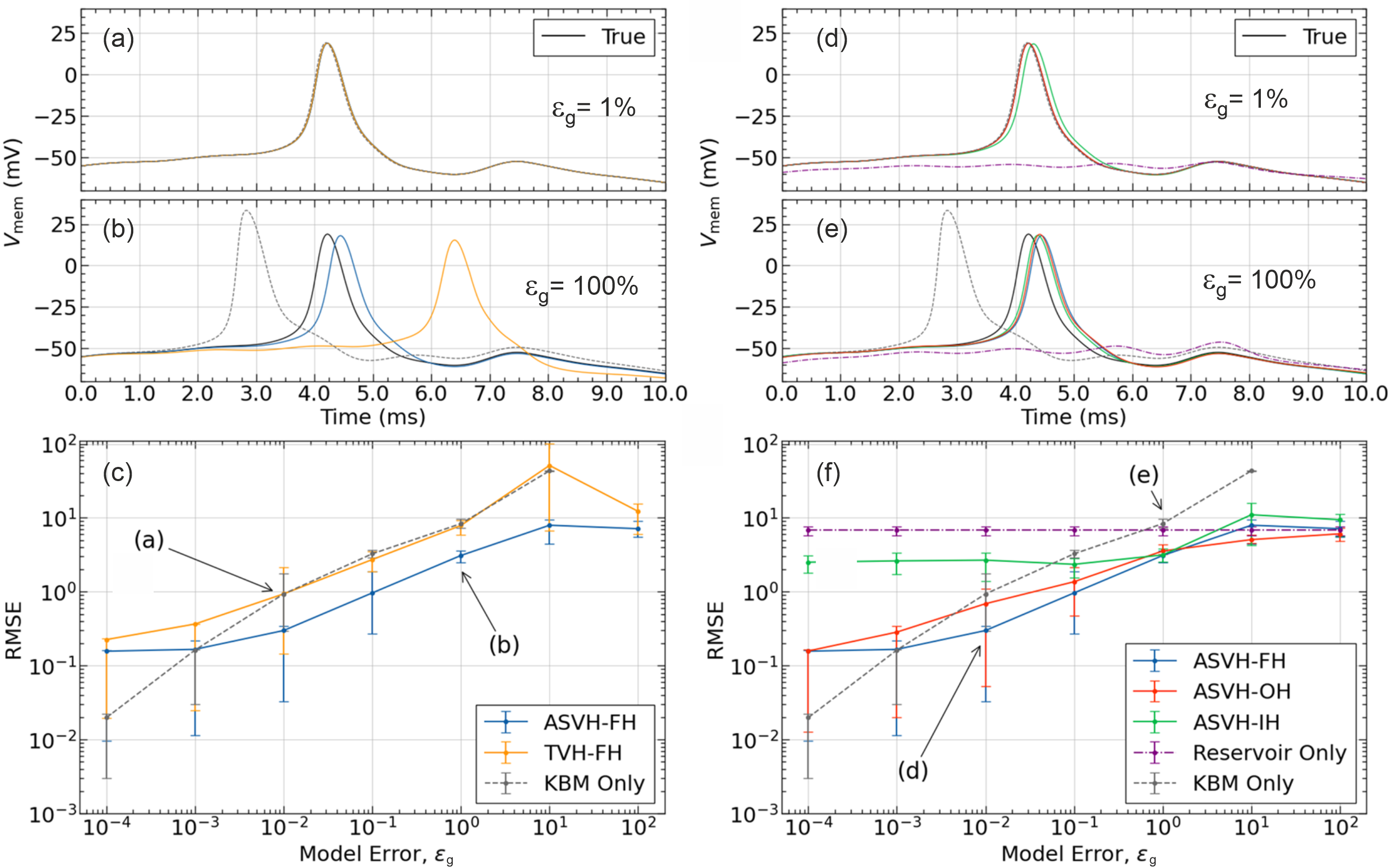}
\caption{\textbf{Surrogate HH-model corrected by ASVH and TVH architectures} \\  Action potential waveform corrected by the TVH-FH (orange line) and ASVH-FH (blue line) architectures for (a) a small model error: $\epsilon_g=1\%$, (b) a larger model error: $\epsilon_g=100\%$.  Action potentials generated by the surrogate HH model alone (dashed line) and the reference action potential (black line) are also shown.  (c) Dependence of the RMSE deviation of the ASVH-FH (blue trace) and TVH-FH (yellow trace) membrane voltage on model error ($\epsilon_g$).  The RMSE deviation of the raw surrogate model (dashed line) is shown for reference.\\
Action potential waveform corrected by the ASVH-IH (green line), ASVH-OH (red line) and ASVH-FH (blue line) configurations for (d) a small model error: $\epsilon_g=1\%$ and (e) a large model error $\epsilon_g=100\%$.  (f) Dependence of the RMSE deviation of the ASVH-IH (green trace), ASVH-OH (red trace), ASVH-FH (blue trace) membrane voltage on model error.  Error bars show the interquartile dispersion of the RMSE computed from 5 different intervals of the predicted membrane voltage, each 1500ms long.  KBM $\equiv$ surrogate HH model.}
\label{fig:fig4}
\end{figure*}

We now report on model error correction efficacy when the reservoir is updated by the observed state variable only (TVH) as opposed to both the observed and unobserved variables (ASVH).   Figs.\ref{fig:fig4}a and \ref{fig:fig4}b show the action potential waveform of a surrogate model (dashed line) corrected by the TVH-FH architecture (yellow trace) and the ASVH-FH architecture (blue trace).  At $\epsilon_g=100\%$, the action potential of the raw surrogate model and TVH-FH stand several standard deviations away from the reference peak (black trace) and in opposite directions.  In contrast, the action potential corrected by ASVH-FH (blue trace) is almost fully re-synchronized with the reference action potential.  In this example, the ASVH-FH architecture is thus much more effective in correcting model error than TVH-FH.

This is evidenced further by calculating the RMSE prediction errors of the ASVH-FH and TVH-FH architectures and plotting their dependence on the magnitude of model error $\epsilon_g$ (Fig.\ref{fig:fig4}c).  Over the entire error range, $0.01\%<\epsilon_g<10^4\%$, ASVH-FH is more accurate than TVH-FH.  The ASVH-FH architecture also successfully improves the predictions of the raw surrogate model when $\epsilon_g>0.1\%$.  At the smallest errors however, the stand-alone surrogate model is more accurate than either ASVH-FH or TVH-FH.  We interpret this as the consequence of reservoir noise which places a lower limit to model error correction by hybrid architectures.  Reservoir noise is an unavoidable consequence of applying regularization to the calculation of output weights.  Overall these results show that the ASVH-FH configuration produces superior error correcting capability over TVH-FH.  This is the configuration adopted for the remainder of the paper.

\subsection*{ASVH-IH, ASVH-OH and ASVH-FH variants}

We now compare the effect of re-injecting the state vector $\mathbf{x}$ in the input layer only (ASVH-IH), the output layer only (ASVH-OH) and both the input and output layers (ASVH-FH).  Figs.\ref{fig:fig4}d and \ref{fig:fig4}e compare the action potentials corrected by these three variants.  All three successfully rectify the action potential delay of the surrogate HH-model (dashed trace).  When model error is large, $\epsilon_g=100\%$, rectification applied by all three variants is equally effective.  The rectified action potentials almost perfectly match the reference action potential (black trace).  When model error is small $\epsilon_g=1\%$, the -OH and -FH variants perform significantly better than -IH.

This is shown in the dependence of the RMSE prediction error on $\epsilon_g$ (Fig.\ref{fig:fig4}f).  The RMSE prediction error of ASVH-IH (green trace) saturates below $\epsilon_g=100\%$ whereas ASVH-OH and -FH continue to become more and more accurate as $\epsilon_g$ decreases until $\epsilon_g=0.01\%$.  These results show that the FH and OH variants are best for correcting model error.  This is because both weight in the HH model output in the calculation of $\widetilde{V}$.  This allows the surrogate HH model to improve predictions of the combined reservoir-model architecture particularly in the range of smaller model errors.  The ASVH-FH is the approach we retain in the remainder of the paper on the basis that injecting the state vector in the input and output layers may offer situational performance benefits over ASVH-OH.

\subsection*{Reservoir size and fraction of nodes updated by the model}

\begin{figure}
\centering
\includegraphics[width=\linewidth]{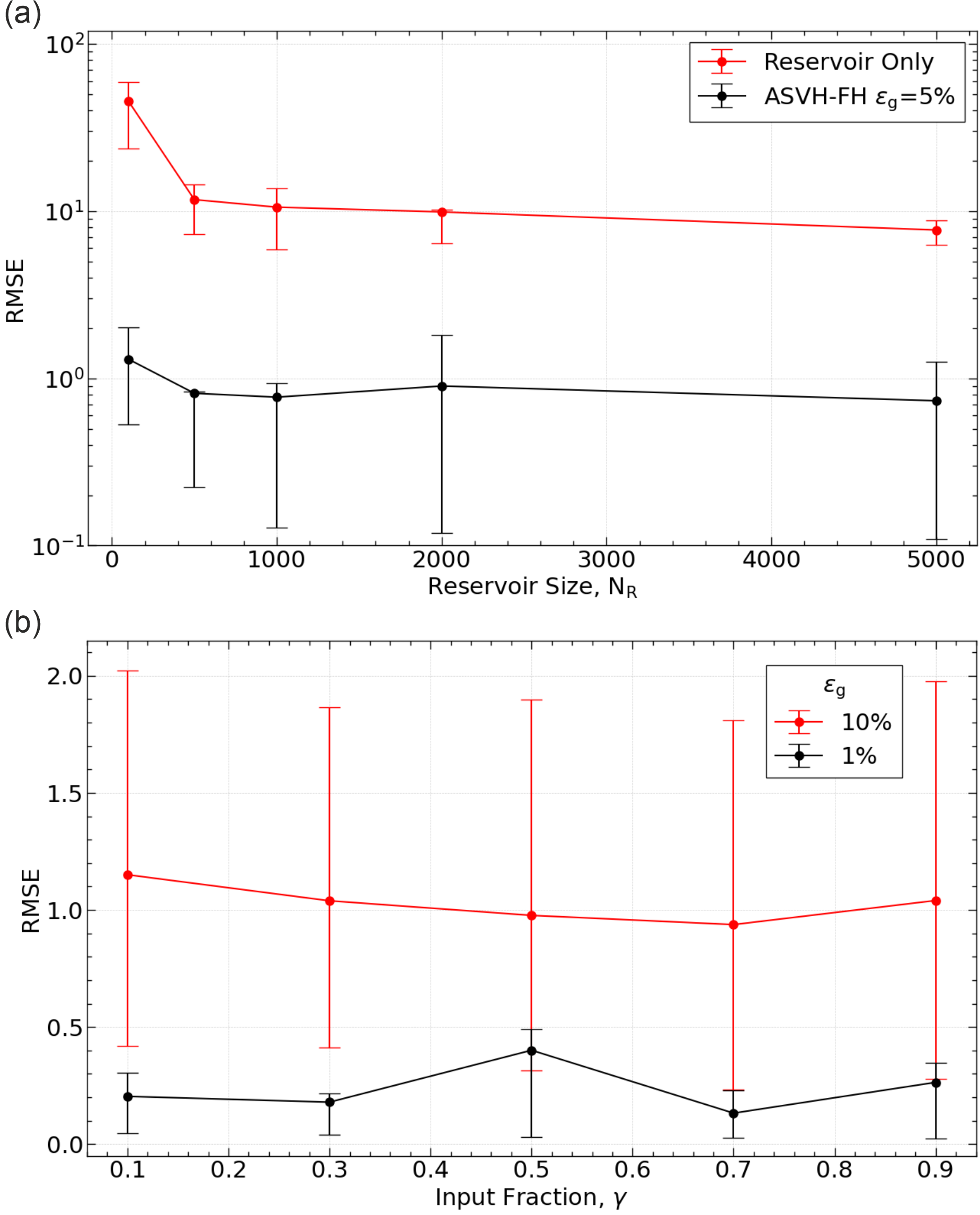}
\caption{\textbf{Optimal reservoir size and ratio of model/training data updating the reservoir} \\
(a) Dependence of RMSE prediction error (ASVH-FH) on the number of nodes $N_R$ in the reservoir (black line).  The corresponding dependence for the reservoir alone is shown for reference (red line).  (b) Dependence of RMSE prediction error (ASVH-FH) on the ratio of reservoir nodes updated by the surrogate HH model ($\mathbf{x}(t)$) over reservoir nodes updated by the training data ($V_{mem}(t)$, $I(t)$).}
\label{fig:fig5}
\end{figure}

The dependence of ASVH-FH prediction accuracy on reservoir size is plotted in Fig.\ref{fig:fig5}a.  The larger the reservoir, the more accurate the predictions.  However RMSE prediction error decreases at a marginal rate once $N_R$ has reached 1000 nodes.  This is why we use $N_R=1000$ as a trade-off between minimizing computation time and maximising prediction accuracy.

We then sought to determine the optimum fraction $\gamma$ of reservoir nodes updated by the model.  Fig.\ref{fig:fig5}b plots the dependence of the ASVL-FH prediction error as $\gamma$ varies between $0.1$ and $0.9$.  The traces corresponding to $\epsilon=1\%$ and $10\%$ revealed no discernible dependence of the RMSE error on $\gamma$.  Therefore we used a value of $\gamma=0.5$ as default.

\subsection*{Correcting different types of model error}

\begin{figure*}
\centering
\includegraphics[width=\linewidth]{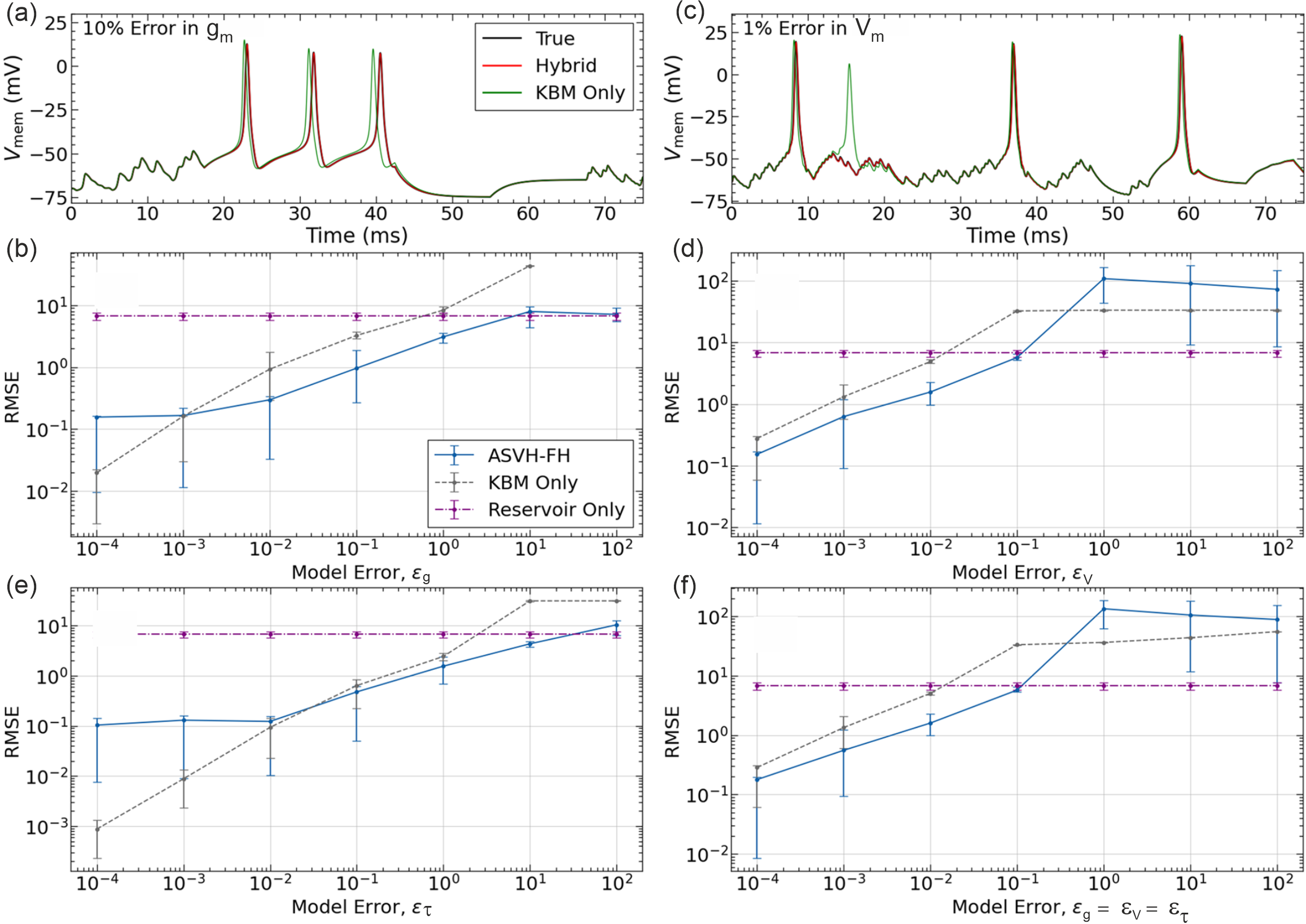}
\caption{\textbf{Correcting different types of model error} \\  (a) Action potentials induced by a 10\% error ($\epsilon_g$) in the sodium channel conductance (green line) and corrected by ASVH-HH (red line).  (b) Dependence of the RMSE prediction error on model error $\epsilon_g$ for the surrogate HH model alone (dashed line) and after correction by ASVH-FH (blue line).  (c) Action potentials induced by a 1\% error ($\epsilon_V$) in sodium activation threshold (green line) and corrected by ASVH-FH (red line).  (d) Same as (b) for $\epsilon_V$ varying.  (e) Same as (b) for the sodium activation time varying by $\epsilon_\tau$ relative to its nominal value.  (f) Same as in (b) for $\epsilon_g$, $\epsilon_v$ and $\epsilon_\tau$ varying all at once.  (a,b) The reference voltage $V_{mem}(t)$ is the black trace.}
\label{fig:fig6}
\end{figure*}

So far, linear model error has been applied by detuning the Na conductance.  We now investigate the efficacy of the ASVH-FH architecture in correcting nonlinear model error caused by an erroneous Na activation threshold or erroneous time constant of the sodium activation gate.  The results are shown in Fig.\ref{fig:fig6}.

For reference, Fig.\ref{fig:fig6}a shows action potentials induced by an $\epsilon=10\%$ model error and after correction by ASVH-FH.  The corrected action potentials (red line) are identical to the reference (black line).  In contrast those of the uncorrected surrogate model (green line) exhibit a significant delay.  The efficacy of model error correction is further validated in Fig.\ref{fig:fig6}b.

Fig.\ref{fig:fig6}c shows the effect of correcting nonlinear error $\epsilon_V=1\%$ with the ASVH-FH architecture.  The uncorrected surrogate HH model (green trace) exhibits an extra action potential at $t=16$ms.  After ASVH-FH correction, the predicted action potentials (red line) match those of the reference time series (black line).  The dependence of the RMSE prediction error of the ASVH-FH architecture was further plotted as a function of $\epsilon_V$ (Fig.\ref{fig:fig6}d).  The ASVH-FH predictions were found to successfully correct the surrogate HH model as long as $\epsilon_V<40\%$.  When $\epsilon_V>40\%$, the stand-alone reservoir is more accurate than either ASVH-FH or the uncorrected surrogate model.

Fig.\ref{fig:fig6}e shows the efficacy of the ASVH-FH architecture in correcting nonlinear error $\epsilon_\tau$.  This type of error is the most severe as it affects gate variables exponentially through Eq.\ref{eq:eq2}.  The ASVH-FH system is found to remain effective in correcting this model error (blue trace) however the gain in prediction accuracy is markedly smaller than in other types of model error ($\epsilon_V$, $\epsilon_g$).  ASVH-FH is most effective at correcting the largest model errors ($\epsilon_\tau>100\%$) when the output weights decouple the reservoir from the surrogate model.

Fig.\ref{fig:fig6}f lastly considers the combined effect of the three types of model errors $\epsilon_g$, $\epsilon_V$ and $\epsilon_\tau$ which are varied all at once.  This plot shows that the AVSH-FH architecture improves surrogate predictions up to model error of $\epsilon_g=\epsilon_V=\epsilon_\tau<40\%$.

These results show that the ASVH-FH approach is effective at improving the predictions of surrogate HH models corrupted by combined errors in linear and nonlinear parameters.

\subsection*{Correcting model error on unobserved state variables: $m$,$n$ and $h$}

The above results have presented the effects of model error correction in the observed state variable $V(t)$.  In this section we examine whether the ASVH-FH architecture is also able to correct the dynamics of the unobserved variables $m(t)$, $h(t)$ and $n(t)$.  A well-known challenge when synchronizing a complex nonlinear systems to a time series~\cite{meliza_estimating_2014,nogaret_automatic_2016,OLeary2015,Kennel1992,Letellier2005,Schumann2016,Aeyels1981} is the multiplicity of solutions creeping in the optimisation problem, and a fortiori when the fitting model is known to be wrong~\cite{Wells2024,nogaret_approaches_2022}.  The ability to successfully synchronize one state variable $V(t)$ to $V_{mem}(t)$ does not in general guarantee that all other state variables will match the dynamics of the reference model.  In this respect, our knowledge of the original HH model will be particularly useful here to validate the error corrected $m(t)$, $h(t)$, $n(t)$ waveforms against the same waveforms in the reference model.

\begin{figure*}
\centering
\includegraphics[width=\linewidth]{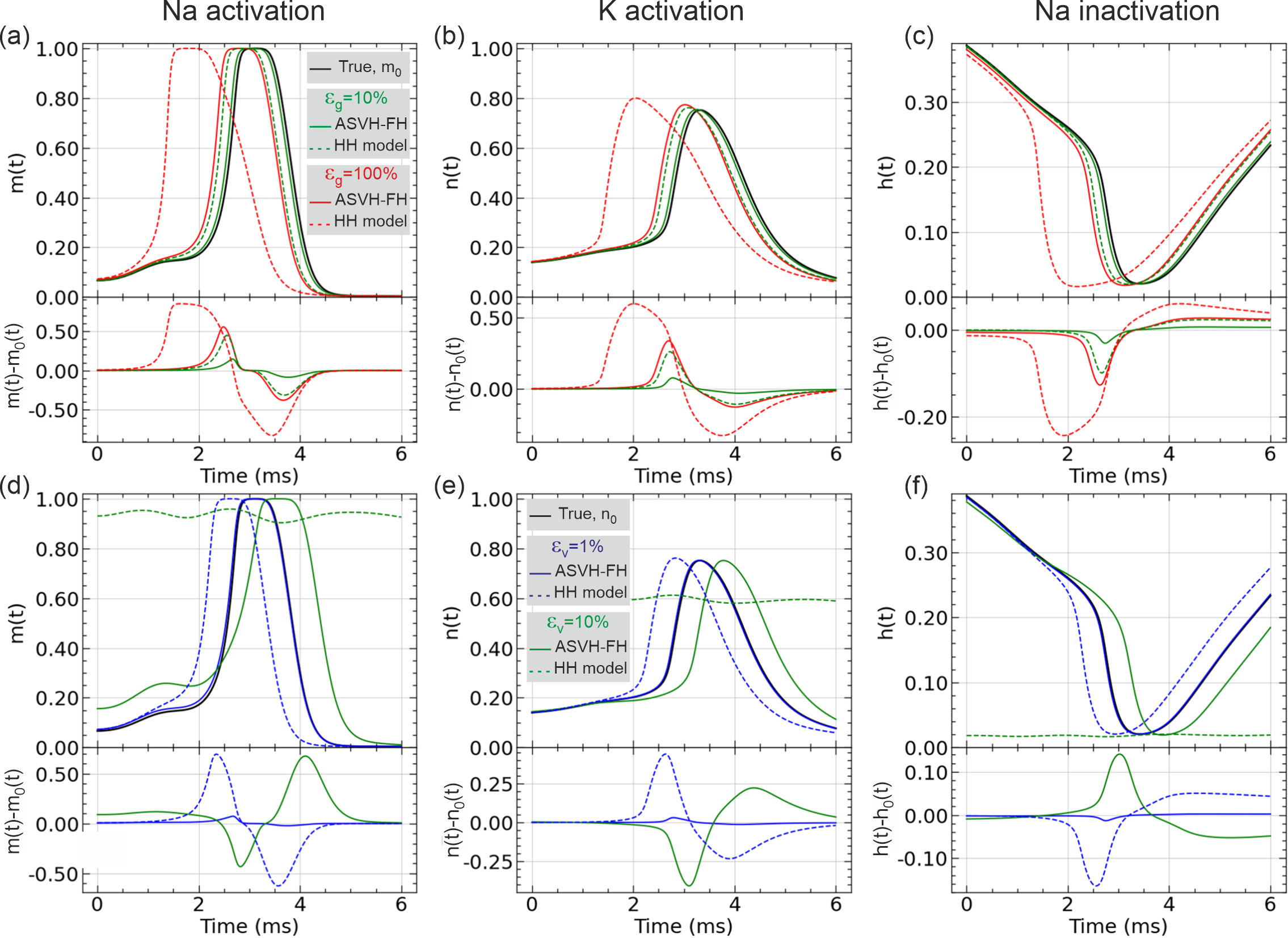}
\caption{\textbf{Correcting error in the unobserved gate variables} \\ Dynamics of ionic gates (a) $m(t)$ (Na activation), (b) $n(t)$ (K activation) and (c) $h(t)$ (Na inactivation) predicted by a surrogate HH model with an erroneous sodium conductance detuned from nominal value by $\epsilon_g=10\%$ (green dashed line) and $100\%$ (red dashed line).  Ion gate dynamics corrected by ASVH-FH for $\epsilon_g=10\%$ (full green line) and $100\%$ (full red lines).  Panels (d-f) plots the $m(t)$, $n(t)$ and $h(t)$ waveforms with $\epsilon_v=1\%$ error (dashed blue line), $10\%$ error (dashed green line) in the Na activation threshold, and after correction with ASVH-FH (full blue line and green lines respectively).}
\label{fig:fig7}
\end{figure*}

Figs.\ref{fig:fig7}a-c show the waveforms of two surrogate models with $10\%$ and $100\%$ error in the sodium conductance and after ASVH-FH correction.  Figs.\ref{fig:fig7}d-e show the same for $1\%$ and $10\%$ error in the sodium activation threshold.  The waveforms describe the gate dynamics during a single action potential.  The magnitude of model error correction can be seen in the uncorrected waveforms (dashed lines) and after ASVH-FH correction (full lines).  In all panels, the error corrected gate variables nearly match the reference waveform (black line).  The largest discrepancy is observed in the $\epsilon_V=10\%$ traces (green lines).  Note in this case that model error is so large that the surrogate HH model does not oscillate anymore (dashed green lines).  The ASVH-FH corrected waveforms (full green lines) have their oscillations restored and their action potential is re-synchronized to the reference action potential (black line) within less than a standard deviation.  This is a remarkable result given that the ASVH-FH architecture received information on the gate dynamics indirectly from time delayed measurements of $V_{mem}$~\cite{Rey2014}.  We recall that Takens' theorem~\cite{Takens1981,Aeyels1981,Sauer1991,Schumann2016} prescribes that a sequence of 9 time delayed measurements $V_{mem}(t-4\Delta t),\dots,V_{mem}(t+4\Delta t)$ is required for constraining the 4-components of the state vector at time $t$.  The inference of unobserved variables therefore places a further requirement on the reservoir retention time to keep track of the observed voltages on a scale of at least $8\Delta t$.  Although we have not systematically studied the effect of the reservoir retention time on the prediction accuracy of the unobserved gate variables, we found that increasing the degree of the adjacency matrix from $D=6$ to $8$ and decreasing the spectral radius from $\rho=1.25$ to $1$ improved ASVH-FH predictions (Table~\ref{tab:tab1}).  The effect of these changes was to increase the retention time of the reservoir~\cite{Ebato2024}.

Overall, the hybrid ASVH-FH architecture has largely recovered the error-free dynamics of both observed and unobserved state variables.  The algorithm has demonstrated its robustness to various types of model error and its relevance to building accurate models from partial information from the reference model.

\section*{Discussion}

This paper has made progress in three directions.  Firstly, we have modified the reservoir architecture to train a driven nonlinear oscillator as opposed to a self-sustaining oscillator such as the Kuramoto-Sivashinsky~\cite{Pathak2018} or the Lorenz system~\cite{duncan_optimizing_2023,pathak_hybrid_2018}.  Secondly, we have inferred the dynamics of unobserved gate variables by synchronizing a surrogate HH model to the reference time series.  The error-free dynamics of the reference model was successfully recovered by coupling the surrogate model to a reservoir.  Thirdly, we investigated different types of model error by detuning linear and nonlinear parameters.

\subsection*{Optimal Reservoir - HH model architecture}

We have shown that a stand-alone reservoir could emulate the membrane voltage dynamics of a neuron driven by external time dependent current.  However predictions were unstable.  Rapid current oscillations or abrupt changes in current amplitude would propel the reservoir into a non-oscillatory state (Fig.\ref{fig:fig1}).  Coupling the reservoir to a conductance-based neuron model (Fig.\ref{fig:fig3}) solved this problem by making predictions sustainable over a longer range.  This ability of hybrid reservoir-model architectures to extend the prediction range was already noted in models of self-sustaining oscillators~\cite{pathak_hybrid_2018,duncan_optimizing_2023}.  Of the 6 possible hybrid architectures: ASVH-FH, ASVH-OH, ASVH-IH, TVH-FH, TVH-OH, TVH-IH we considered in Fig.\ref{fig:fig3}, the configuration that best corrects model error is ASVH-FH.  This is because the reservoir was updated with the complete information from the state vector (ASVH) and the ridge regression of output weights incorporated both the complete information from the state vector and from the reservoir (-FH).

The optimal value of reservoir hyperparameters was determined empirically in Table~\ref{tab:tab1}. The fraction of reservoir nodes updated by the model could be set to $\gamma=50\%$ in all simulations as prediction accuracy was independent of this parameter in the 0.1-0.9 range (Fig.~\ref{fig:fig5}b).  This is because the connection to the input layer and its fixed weights (Fig.~\ref{fig:fig3}) is of lesser importance than the connection to the variable weight output layer.  This result concurs with Duncan and Räth's observations~\cite{duncan_optimizing_2023} for self-sustained oscillators.  The optimum number of reservoir nodes was $N_R=1000$.  This is a tradeoff between accuracy and computational cost as largest reservoirs do not significantly increase prediction accuracy.

Hyperparameters $D$ and $\rho$ were assigned different values in the stand alone Reservoir and the Reservoir-HH model (Table~\ref{tab:tab1}).  It was empirically determined that the reservoir of the hybrid architecture needed a greater memory retention capacity.  This was realized by increasing the delay capacity of the neural network.  In practice this meant reducing the spectral radius $\rho$ and increasing the adjacency degree $D$ to reduce signal transfer efficiency within the network while conserving a good fitting ability.  The need for greater memory capacity was called for transferring information from time delayed measurements of $V_{mem}(t)$ to the unobserved state variables of the HH model at time $t$.  The HH model has three unobserved variables.  According to embedding theory~\cite{Takens1981,Aeyels1981,Letellier2005}, $2n+1$ time delayed voltage measurements are required to determine a $n$-component state vector at time $t$.  Hence the greater the complexity of a neuron model, with additional ion channels and gate variables, the greater the network retention time needs to be.  For the stand-alone reservoir fitting a single variable, a smaller retention time was sufficient (Fig.\ref{fig:fig1}a).

A long reservoir retention time is paradoxically detrimental to the accuracy of predictions when the reservoir has to respond to fast stimulation.  Following a current-pulse an action potential is normally followed by the membrane voltage returning to its resting value (Fig.\ref{fig:fig1}, black trace) and the gate variables returning to their respective resting values (Fig.~\ref{fig:fig7}).   In contrast, even a well-trained reservoir system will fail to precisely return to its rest value (Figs.~\ref{fig:fig4}d,e, dashed line).  It will only do so after several milliseconds.  Residual error in the membrane voltage will therefore accumulate when a series of current pulses is applied on a time scale shorter than the network retention time.  This is the reason why rapid current pulses of large amplitude often send the reservoir into saturation as seen in Fig.\ref{fig:fig1}a.  Coupling a HH model to the reservoir in the ASVH-FH architecture has the benefit of correcting the slow dynamics of the network during integration of the current stimulus.  Even erroneous HH models are useful in providing the speed needed to respond to fast forcing.  This is why we see that all ASVH-FH predictions, including a surrogate or reference model (Figs.\ref{fig:fig1},\ref{fig:fig4},\ref{fig:fig5},\ref{fig:fig7}), remain remarkably stable and accurate over long time intervals.

\subsection*{Model error correction}

Within the ASVH-FH architecture, the HH model and the reservoir were found to effectively compensate each other to minimise prediction error.  The output weights are trained to prioritize the contribution of one rather than the other to best fit the reference voltage time series.  However the Tikhonov regularisation term, $\beta$, employed in the training process to prevent overfitting, also allows the reservoir output to deviate from true value.  This deviation repeated in discrete time over the training window generates a base noise that limits the performance of model error correction.   This is evident at low model error: $\epsilon_g<0.1\%$, $\epsilon_\tau<1\%$, where the surrogate HH model happens to be more accurate than the reservoir-surrogate model hybrid (Figs.\ref{fig:fig6}b,\ref{fig:fig6}e).  The reservoir component fails to learn and correct model error once it has become smaller than the r.m.s. noise.  While this effect is unlikely to meaningfully impact practical applications due to the low model error required for the effect to be observed, it remains an important consideration when investigating near-perfect models.

In the opposite case where model error is so large as to generate grossly inaccurate predictions, the output weights decouple the hybrid architecture output $\widetilde{V}$ from the surrogate model via the output layer (Fig.~\ref{fig:fig3}).  This is necessary as the reservoir component is constantly updated by the model via untrained input weights.  In this regime the hybrid reservoir-model architectures behaves increasingly like the stand-alone reservoir as seen in the saturation of ASVH-FH and TVH-FH at $\epsilon_g>1000\%$ (Fig.\ref{fig:fig4}c), and ASVH-FH at $\epsilon_V>100\%$ (Fig.\ref{fig:fig6}d).

\subsection*{Parameter error correction}
The hybrid ASVH-FH approach may be extended to estimating error-free parameters of biological neurons.  Model parameters are unobserved state variables whose time derivative is equal to zero.  It would therefore suffice to include in the surrogate HH model additional rate equations similar to Eq.\ref{eq:eq2} but with a rate of change equal to zero.

\subsection*{Model error correction in actual neurons and neural networks}
The error correcting approach of presented here was applied in the context of the Hodgkin-Huxley model.  The method can easily be adapted to multi-channel, multi-compartment conductance models describing actual neurons and networks.  The ASVH-FH method can also be adapted to be include multiple forcing inputs, such as multiple current protocols and/or optogenetic stimulation; multiple reference time series voltages, such as multi-electrode measurements and/or calcium concentration readings.  These multiple input components would be allocated to different segments of the input layer.  All model state variables would be input to both $\mathbf{W}_{\mathrm{in}}$ and $\mathbf{W}_{\mathrm{out}}$ generalizing the ASVH-FH approach.

\section*{Conclusion}

The hybrid reservoir-model approach introduced by Pathak $\emph{et al.}$~\cite{pathak_hybrid_2018} for self-sustaining oscillators has been extended to forecasting the state of an externally driven dynamical system.  The reservoir was found to effectively correct the effects of model error on predictions when incorporating a surrogate model into the hybrid architecture.  The hybrid reservoir-surrogate HH model architecture largely predicted the error-free waveforms of observed and unobserved variables in the original HH model used to generate the training data.  Several trials were run to determine the optimal configuration of the hybrid reservoir-model system. It was found that the best performance was achieved by injecting all state variables of the embedded model in both the input and output layers of the reservoir rather than a subset of these state variables.  The hybrid system was also able to correct model error over a wide error range, when parameters of the surrogate model were detuned by up to $1000\%$ from their reference value.  It was noted that intrinsic reservoir noise places a lower limit to the model error that can be corrected.  We have discussed the importance of increasing the lag time of the reservoir for training systems with a large number of unobserved variables.  Conversely we have seen that a large reservoir lag time can also limit the ability of a stand alone reservoir to respond to fast forcing oscillations.  However coupling the reservoir to a HH model was found to considerably improve the stability and accuracy of predictions.  The model error correction developed here constitutes significant progress towards extracting reliable information from biological data when only approximate models are available.

\section*{Methods}

Parameters of the reference Hodgkin-Huxley model: \\

\noindent \textit{Sodium channel}: $g_{Na}$=69 mS.cm$^{-2}$, $E_{Na}$=41mV, $V_m$=-39.92 mV, $dV_m$=10 mV, $\tau_{0,m}$=0143 ms, $\epsilon_m$= 0.1 ms, $dV_{t,m}$=23 mV, $V_h$=-65.37 mV, $dV_h$=-17.65 mV, $\tau_{0,h}$=0.701 ms, $\epsilon_h$= 12.9 ms, $dV_{t,h}$=27.22 mV.

\noindent \textit{Potassium channel}: $g_{K}$=6.9 mS.cm$^{-2}$, $E_{K}$=-100mV, $V_n$=-34.58 mV, $dV_n$=22.17 mV, $\tau_{0,n}$=1.291 ms, $\epsilon_n$= 4.314 ms, $dV_{t,n}$=23.58 mV.

\noindent \textit{Leak channel}: $g_{L}$=0.165 mS.cm$^{-2}$, $E_{L}$=42mV, $C=$1.0$\mu$F.cm$^{-2}$.

\bibliography{Bib_Neuromodcorr}

\section*{Author contributions statement}

JDT demonstrated the initial the proof of concept of predicting neuronal oscillations with a reservoir driven by a time dependent current.  AN conceived the approach of correcting model error in neuron-based conductance models.  IW performed the simulations of hybrid architectures and generated all proof of concept data.  IW and AN wrote the manuscript.  All authors reviewed the manuscript.

\section*{Additional information}

\noindent \textbf{Competing interests:} The authors declare no competing interests.

\noindent \textbf{Data availability statement:} The data and code that support the findings of this study are available within the article and from the authors.

\end{document}